\documentclass[12pt]{article}

\usepackage[margin=1in]{geometry} 
\usepackage{amsmath,amsthm,amssymb}
\usepackage{textcomp}
\usepackage[dvips,final, pdftex]{graphicx}         		
\usepackage[greek, english]{babel}                			
\usepackage[hang,bf]{caption}              			
\usepackage{rotating}                     		 		
\usepackage{subfigure}
\usepackage{units}
\usepackage{fancyhdr}
\usepackage{lscape}
\usepackage{amsfonts}
\usepackage{float}
\usepackage{graphicx}
\usepackage{longtable}
\usepackage{fancybox}
\usepackage[hidelinks]{hyperref}
\usepackage{here}
\usepackage{setspace} 
\usepackage{units} 
\usepackage{pdfpages}
\usepackage{lscape} 		
\usepackage{pdflscape}

\usepackage{upgreek}

\usepackage{listings}

\begin{document}
\bibliographystyle{elsarticle-num}	
	
\title{A review of partial slip solutions for contacts represented by half-planes including bulk tension and moments}

\author{H. Andresen$^{\,\text{a,}}$\footnote{Corresponding author: \textit{Tel}.: +44 1865 273811; \newline \indent \indent \textit{E-mail address}: hendrik.andresen@eng.ox.ac.uk (H. Andresen).}, D.A. Hills$^{\,\text{a}}$$\,\,$\\ \\
	\scriptsize{$^{\text{a}}$ Department of Engineering Science, University of Oxford, Parks Road, OX1 3PJ Oxford, United Kingdom}}
\date{}
\maketitle

\begin{center}
	\line(1,0){470}
\end{center}
\begin{abstract}
Solution procedures for establishing the regimes of stick and slip
for frictional contacts capable of idealisation within the framework of half-plane
elasticity and subject to complex loading regimes are reviewed, starting with the well known fundamental
problems and going on to more complicated ones. These include problems
where the normal load, shear force, applied moment and differential
remote tensions all vary with time. Transient and steady state solutions
are discussed.\\

\end{abstract}
\begin{center}
	\line(1,0){470}
\end{center}

\section{Introduction}
\begin{figure}[t]
	\centering
	\includegraphics[scale=0.75, trim= 0 0 0 0, clip]{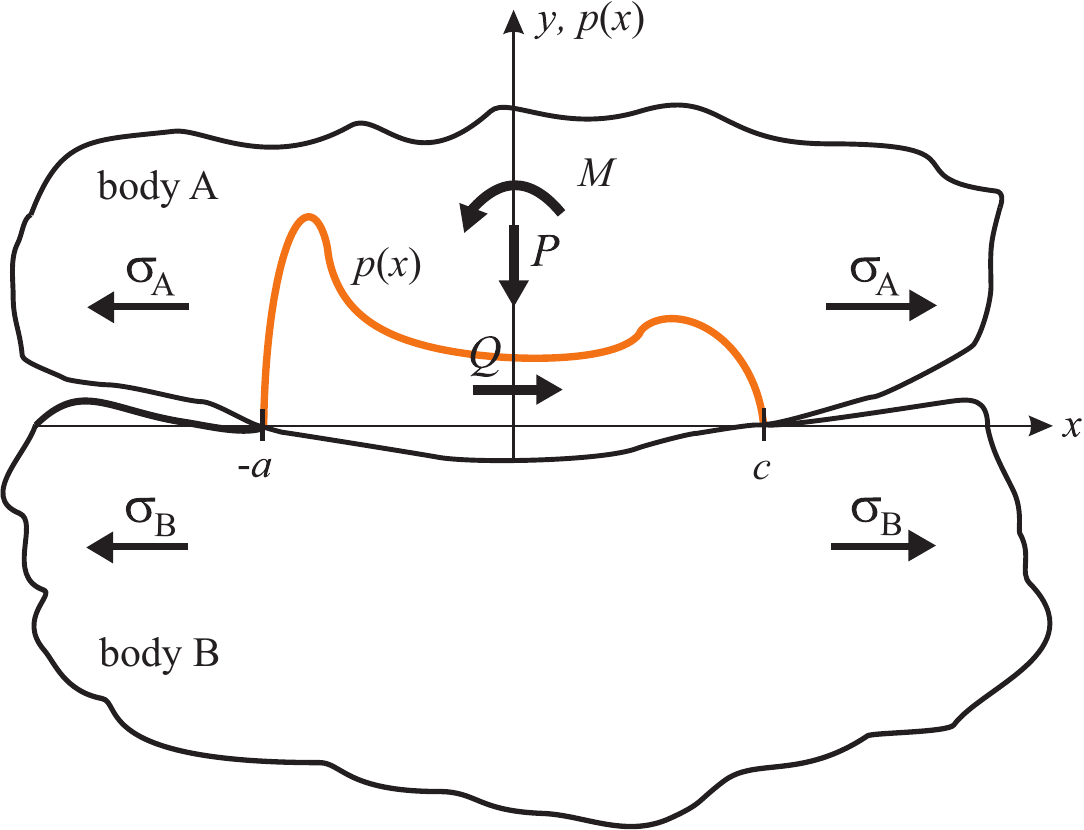}
	\caption{Generic half-plane contact subject to normal load, moment, shear load and bulk tension.}
	\label{fig:generic_half_plane}	
\end{figure}

\hspace{0.4cm}Static joints arise in many mechanical assemblies, and whilst the
majority conform, a number of heavily loaded contacts have rounded
edges. Examples include swashplate assemblies in helicopter rotor
heads, dovetail roots in gas turbine fan blades, and in the locking
segments used to fasten risers to well heads in subsea installations.
These contacts invariably suffer complicated load histories and, as
they are `incomplete' in character the contact pressure falls
smoothly to zero at the edges, so that they are prone to local slip. In fact,
unless very special loading trajectories are followed, there will
inevitably be some periods of slip with attendant fretting damage.
Although some incomplete contacts which nearly conform must be modelled
using elasticity formulations for the particular domain shape corresponding
to their form (e.g. disk, anti-disk) the majority may adequately be
represented using half-plane formulations \cite{Barber_2018}. The `core' of the problem
is represented by the idealised half-plane problem, Figure \ref{fig:generic_half_plane}, and
in this we can identify five load quantities which are relevant to the
fretting fatigue problem, and all of which may be functions of time. There are; the normal load, $P$, shear force, $Q$, moment, $M$, and differential bulk tension arising from tensions developing in each body, $\sigma = \sigma_{A} - \sigma_{B}$.
If the contacting components are made from the same material the contacts
are `uncoupled' in character so that the normal contact problem may
be solved first and then attention given to the effects of shear tractions
which will have no influence on the normal solution.

Here, we will first review the historical solutions where the normal load is held constant, and then examine in more detail
recent solutions which permit both more general forms of loading to be handled (including a rocking moment and differential bulk tension) and more general loading histories to be tracked out. The majority
of solutions employ a method in which the shear traction distribution
is viewed as the sum of that due to sliding, together with a corrective
term, but progress has also been made with an alternative form in
which a correction is made to the fully-stuck solution.

\section{Constant Normal Load Problems}

\hspace{0.4cm}The earliest and best known normal contact
solution was that found by Hertz, i.e. where the contacting bodies
have second order (strictly parabolic but usual interpreted as circular
arc) profiles \cite{Hertz_1881}, so it is natural that the first partial
slip contact solutions were associated with the same geometry. The
first solution, for a monotonically increasing shear force, was found by
Cattaneo \cite{Cattaneo_1938}, and, apparently unaware of this solution, Mindlin
\cite{Mindlin_1949} developed the same solution and went on to look at unloading
and reloading problems \cite{Mindlin_1951}, \cite{Mindlin_1953}. These were the only significant
solutions for some time, and then Nowell and Hills \cite{Hills_1987} looked
at what happened when a bulk tension was simultaneously exerted in
one body as the shear force was gradually increased. Note that, while
the application of an increasing shear force induces slip zones of
the same sign at each edge of the contact, a tension induces slip
of opposite sign and hence, when the two are exerted simultaneously,
one slip zone will be bigger relative to the `no bulk tension' case
whilst the other will be smaller and, in the case of large tension
may actually slip in the opposite direction. The next breakthrough came with the near simultaneous discovery by J\"ager \cite{Jaeger_1997} and Ciavarella \cite{Ciavarella_1998} that, just as the `corrective' shear traction was a
scaled form of the sliding shear traction for the Hertz case, the
same geometric similarity applies whatever the form of the contact, and it worthwhile understanding the
proof of this principle.

\subsection{J\"ager-Ciavarella Principle}

\hspace{0.4cm}It is assumed that the reader is familiar with the basic integral
equations relating surface slope and surface strains to the surface
tractions present on the surface of a half-plane. For a full explanation of these and their derivation
see \cite{Barber_2010}. Although it is not a necessary condition
in the first part of the calculation, it is very much easier to
follow if we assume that the contact is symmetrical in $x$. This
ensures that the stick zone is centrally positioned, and only its
\textit{extent} needs to be found.  If the
relative profile $g(x)$ of two elastically similar half-planes, each subject to a pressure $p(x)$ over the contact length {[}$-a\quad a${]}, is such that their relative
slope, $\mathrm{d}g/\mathrm{d}x$, the following integral equation defines the implied contact pressure
\begin{align}\label{SIE}
\frac{E}{1-\nu^{2}}\frac{\mathrm{d}g}{\mathrm{d}x}=\frac{4}{\pi}\int_{-a}^{a}\frac{p(\xi)\mathrm{d}\xi}{x-\xi},\quad-a\leq x\leq a \text{,}
\end{align}
where $E$ is Young's modulus and $\nu$ is Poisson's ratio and plane strain conditions obtain. For details of the inversion see Barber \cite{Barber_2010} or Hills et al. \cite{Hills_1993}.
When once the contact pressure is known the size of the contact may
be found from overall normal equilibrium, that is
\begin{align}
P=\int_{-a}^{a}p(x)\,\mathrm{d}x \text{.}
\end{align}

As the contact is formed surface particles will displace laterally
but, because the contact pressure is mutual and the materials are
elastically similar, the displacement is the same in each body so
that here is no difference in surface strain, and no shear tractions
arise. A vanishingly small coefficient of friction is sufficient to
inhibit all slip. Note that this would not be the case if a shear
force were gradually exerted as the normal load was increased, see $\mathsection3$.

Here, we subsequently apply a monotonically increasing shear force so that slip starts at the edges where the contact pressure goes to zero. Suppose that the stick region has contracted to the interval {[}$-b\qquad b${]}. For stick to be maintained in this region the differential strains parallel with the surface must vanish so that
\begin{align} 
\Delta \varepsilon_{xx}=0,\quad-b\leq x\leq b  \text{.}
\end{align} 

\begin{figure}[t]
	\centering
	\includegraphics[scale=0.4, trim=0 0 0 0, clip]{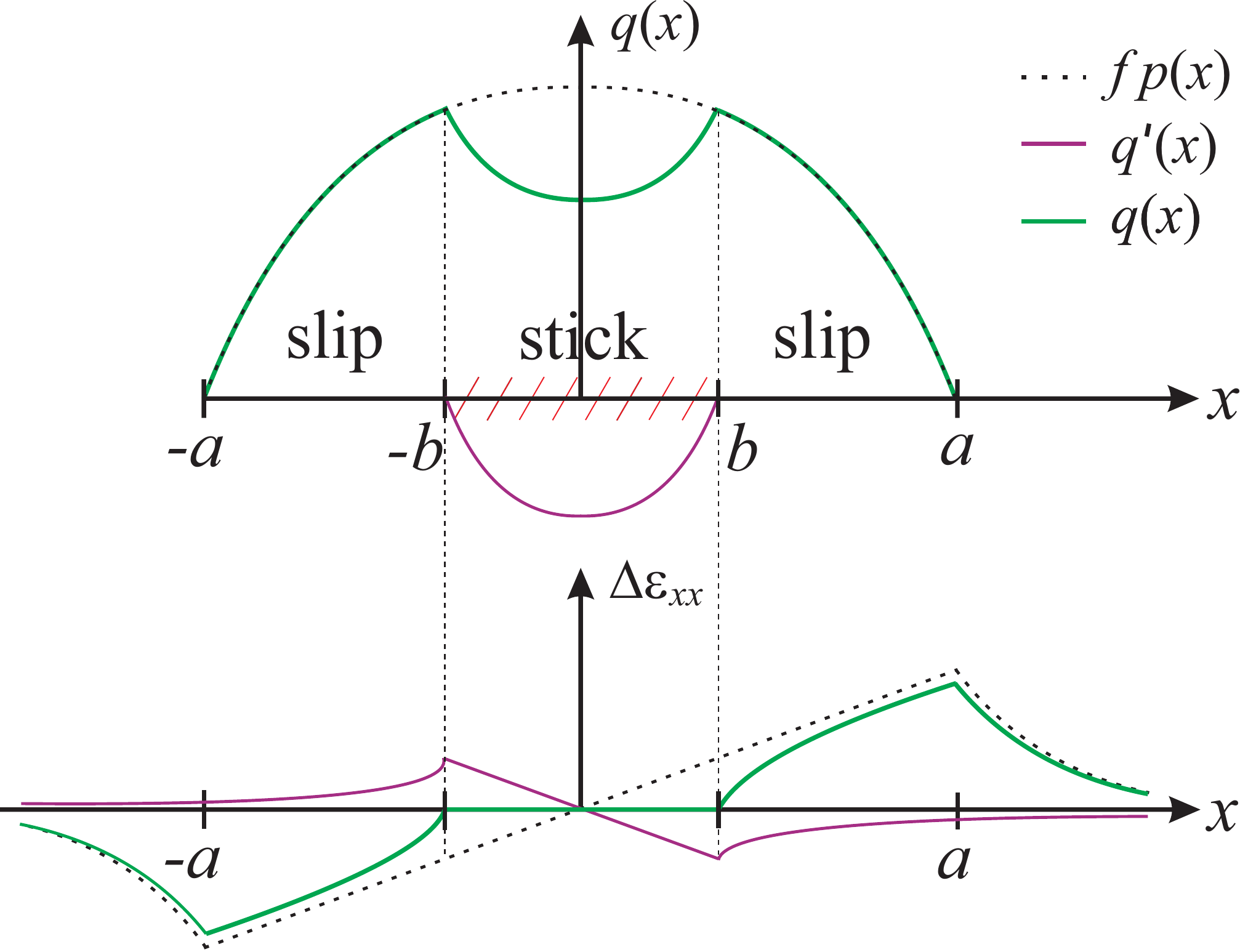}
	\caption{Illustration of shear tractions and relative surface strains for a Hertzian contact in partial slip.}
	\label{CJ_plot}
\end{figure}

Figure \ref{CJ_plot} shows the shear tractions, $q(x)$, and relative surface strains, $\Delta\varepsilon_{xx}$.
The shear traction, $q(x)$, is limited by the coefficient
of friction in the outer slip regions
\begin{align}
q(x)=fp(x) \text{,}\quad b<\left|x\right|<a  \text{,}
\end{align}
where $f$ is the coefficient of friction. The relative strain, $\Delta\varepsilon_{xx}(x)$,
associated with a shear traction $q(x)$ over the interval {[}$-a\qquad a${]}
is given by
\begin{align}
\Delta\varepsilon_{xx}=\frac{4\left(1-\nu^{2}\right)}{\pi E}\int_{-a}^{a}\frac{q(\xi)d\xi}{x-\xi}  \text{,}
\end{align}
and here we write the shear traction as the sum of a contribution
over the whole contact, limited by friction, together with an as yet
unknown second `corrective' term, $q'(x)$, over the stick interval,
giving
\begin{align}
\frac{E}{1-\nu^{2}}\,\Delta\varepsilon_{xx}=\frac{4f}{\pi}\int_{-a}^{a}\frac{p(\xi)\mathrm{d}\xi}{x-\xi}+\frac{4}{\pi}\int_{-b}^{b}\frac{q'(\xi)\mathrm{d}\xi}{x-\xi} \text{.}
\end{align}

But we know that $\Delta\varepsilon_{xx}=0$ within the interval $-b\leq x\leq b$,
so that 
\begin{align}
0=\frac{4f}{\pi}\int_{-a}^{a}\frac{p(\xi)d\xi}{x-\xi}+\frac{4}{\pi}\int_{-b}^{b}\frac{q'(\xi)d\xi}{x-\xi},\qquad-b\leq x\leq b \text{.}
\end{align}

Now compare the first integral with that arising in the solution
for the normal problem, given in equation \eqref{SIE}. Note also that $-a<-b<x<b<a$ which means that
we can re-write the integral equation for the corrective term in the form
\begin{align}
-f\frac{E}{1-\nu^{2}}\frac{\mathrm{d}g}{\mathrm{d}x}=\frac{4}{\pi}\int_{-b}^{b}\frac{q'(\xi)\mathrm{d}\xi}{x-\xi},\qquad-b\leq x\leq b \text{.}
\end{align}

Now, we see that the integral equation for the corrective term and
the integral equation for a normal contact are very similar in form, differing only in
the interval over which they are evaluated. It follows immediately
that the form of the corrective shear traction must be similar to
that for the pressure distribution, but appropriate to a lower load,
which we denote $Q'$. From equilibrium considerations we see that
$Q'=\int_{-b}^{b}q'(x)\mathrm{d}x$, and further 
\begin{align}
Q=fP-Q'\text{.}
\end{align}

\begin{figure}[t]
	\centering
	\includegraphics[scale=0.65, trim=0 0 0 0, clip]{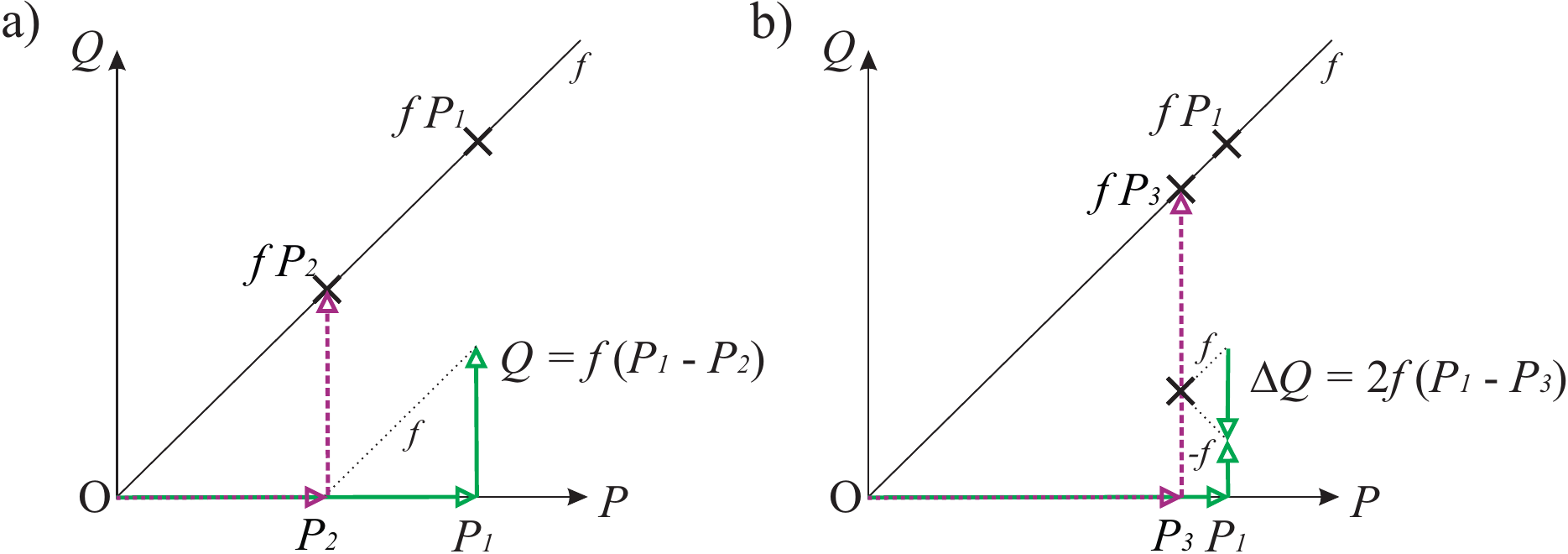}
	\caption{J\"ager-Ciavarella theorem in contracted notation in a $P$-$Q$ plot for a) sequential loading and b) periodic loading.}
	\label{fig:P1_P2_Qprime}
\end{figure}

This is the very simple result needed. We will also be employing it,
in extended forms, for more complicated problems in a later section.
To do this, it is helpful to re-cast the problem in a contracted notation
developed by Barber \cite{Barber_2011}. Suppose we denote the contact pressure distribution,
for whatever geometry of problem we have under consideration, and
when the applied load is $P_{1},$ by $p(x,P_{1})$. Then, in the sequential
loading problem just described, the partial slip shear traction distribution might
be written as 
\begin{align}
q(x)=f\left[p(x,P_{1})-p(x,P_{2})\right]  \text{,}
\end{align}
where the connection with the previous notation is $P_{1}\equiv P,P_{2}=Q'/f$.
This loading history is conveniently displayed on a $P$-$Q$ plot
where time tracks out a loading trajectory, Figure \ref{fig:P1_P2_Qprime} a), and it should
be noted that point $2$ is a point on the previous loading trajectory, here given by
\begin{align}
f=\frac{Q}{P_{1}-P_{2}}  \text{,}
\end{align}
and the stick zone half-width, $b$, is given by $a(P_2)$.
However, this tells only part of the story because, in the majority
of problems, the shear force varies periodically. Upon an infinitesimal
reduction in shear force stick ensues everywhere, and then a further
reduction leads to zones of reverse slip developing. For a full description
of the unloading problem see \cite{Mindlin_1951} and  \cite{Barber_2011}, but let us suppose that the greatest reduction in shear force from its maximum value is $\Delta Q$, which therefore denotes
its range. Then the steady state sick zone half width, $m$, is $a(P_{3})$,
where
\begin{align}
f=\frac{\Delta Q}{2(P_{1}-P_{3})} \text{,}
\end{align}
see Figure \ref{fig:P1_P2_Qprime} b). So, the steady state (or `permanent') stick zone is
greater than the initial one, $b(P_{2})$, and this represents the
phenomenon of `shaking down', that is the self-development of locked
in slip displacements (and hence shear tractions) which tends to reduce
the tendency to slip. We note that the permanent stick zone size depends
on the range of shear stress only and is independent of its mean value. 

When a bulk stress is present varying in phase with the shear load, the position of the permanent stick zone, at least, will be affected. It has been shown by Andresen \cite{Andresen_2019_2} that for a Hertzian contact subject to a shear force together with moderate bulk tension two explicit equations for extent, $d$ and eccentricity, $e$, of the permanent stick zone can be established
\begin{align}\label{eq22}
d^{2}=a^{2}-\frac{A R \text{$\Delta $Q}}{\pi  f}\;\text{,}
\end{align}
and 
\begin{align}\label{eq21}
e=-\frac{A R\Delta \sigma }{8 f} \;\text{.}
\end{align}

In $\mathsection5$ we will address in-phase steady-state solutions on a more general level.

\section{Varying Normal Load Problems}

\begin{figure}[b!]
	\centering
	\includegraphics[scale=0.35, trim=0 0 0 0, clip]{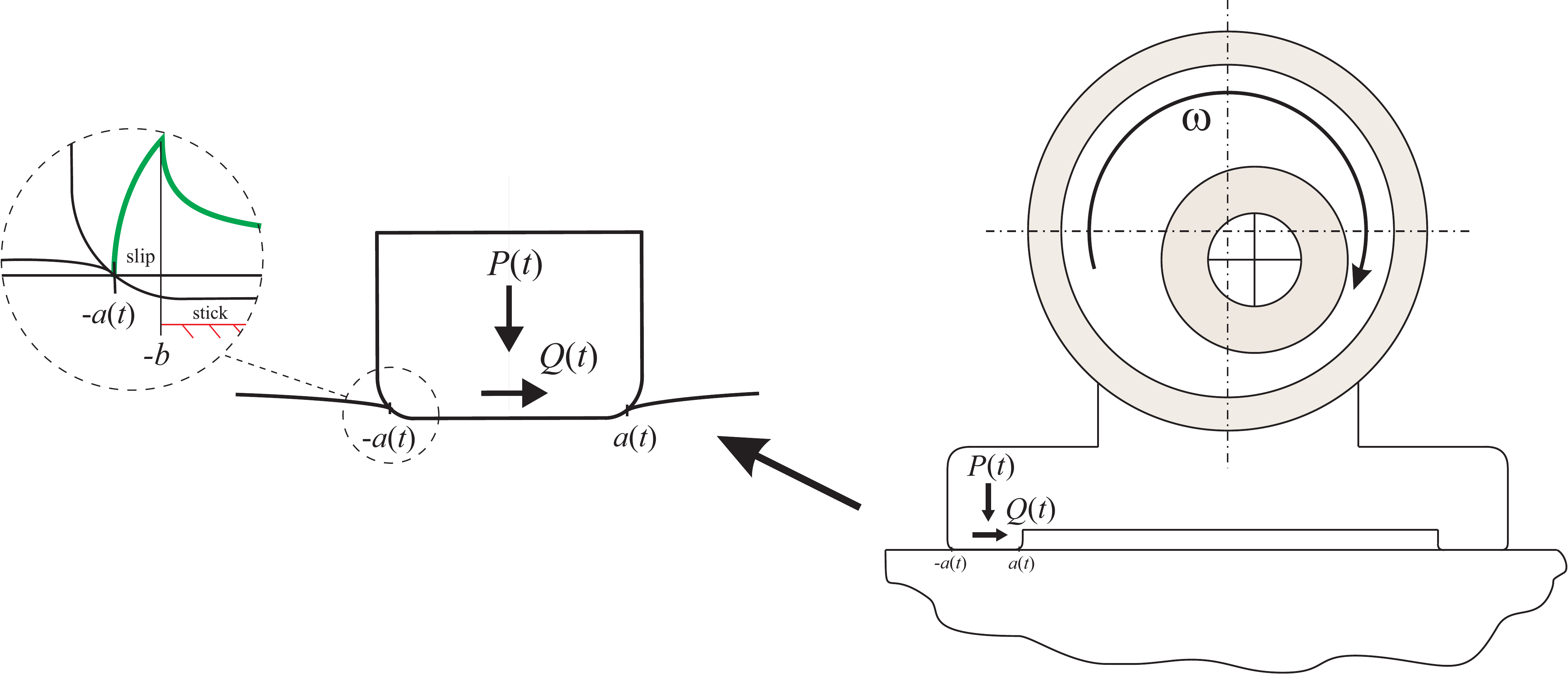}
	\caption{Motor with eccentric shaft oscillating harmonically and half-plane idealisation of its supporting contacts.}
	\label{fig:motor_eccentricity}
\end{figure}

\hspace{0.4cm}Suppose that we have a motor, as depicted in Figure \ref{fig:motor_eccentricity}, whose shaft carries an eccentric weight, and that the motor is supported on contacts which may be idealised using half-plane principles. 
As the shaft rotates both the normal
and shear forces on the contact will change, so that a loop is tracked
out in $P$-$Q$ space of the form
\begin{align}
P(t)=P_{0}+P_{1}\sin\left(\omega t\right)  \text{,}
\end{align}
\begin{align}
Q(t)=Q_{0}+Q_{1}\sin\left(\omega t+\phi\right) \text{,}
\end{align}
where $\omega$ is the angular speed of the shaft and in the example
given the phase shift, $\phi$ is $\pi/2$. $P_0$ and $Q_0$ are the mean values of the load and $P_1$ and $Q_1$ are the respective normal and shear load amplitudes. Major progress in solving problems
of this type, and where the intention was to track out the full behaviour
as a function of time, was made in \cite{Barber_2011}. As before, the stick zone was symmetrically positioned and
only its extent as a function of time was to be found.

When the normal load on a half-plane contact also varies with time,
it is no longer the case that there must, necessarily, be slip at
each time step. If the contact currently has a half-width, $a$, and
the load is increased by $\Delta P$, the contact pressure will change
by
\begin{align}
\Delta p(x)=\frac{\Delta P}{\pi\sqrt{a^{2}-x^{2}}}  \text{.}
\end{align}

Similarly, if the shear force is increased by $\Delta Q$ the shear
traction distribution will change by
\begin{align}
\Delta q(x)=\frac{\Delta Q}{\pi\sqrt{a^{2}-x^{2}}} \text{,}
\end{align}
so that the contact will remain adhered provided that 
\begin{align}\label{fully_adhered_cond}
\frac{\left|\Delta Q\right|}{\Delta P}<f \text{.}
\end{align}

\begin{figure}[t!]
	\centering
	\includegraphics[scale=0.7, trim=0 0 0 0, clip]{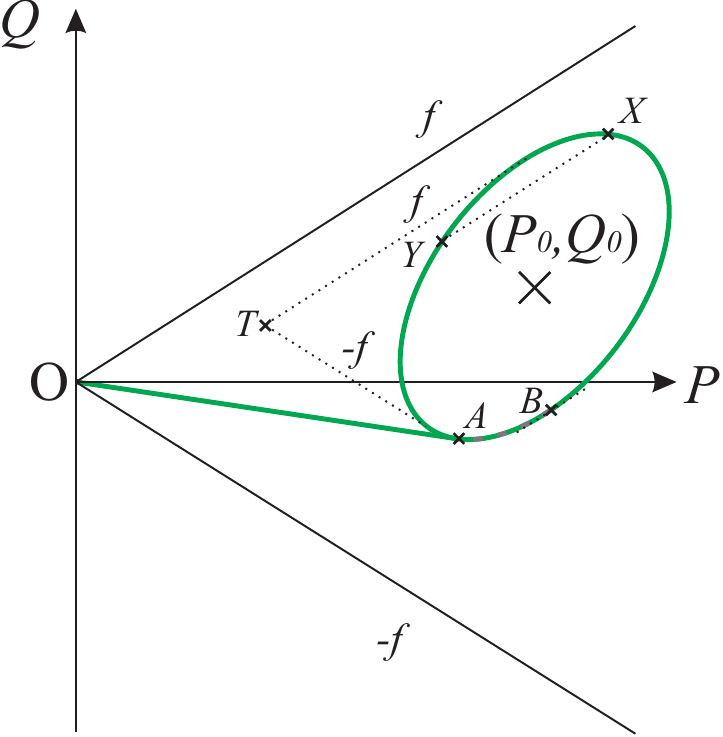}
	\caption{Illustration of a generic loading trajectory in $P$-$Q$ space with point $T$ constructed.}
	\label{fig:PQSpace}
\end{figure}

This result follows directly from the analogy between the Green's
functions for normal and shear loading. The question we now ask is
`How does the shear traction change when we move between two points
($A,B$) on the loading trajectory when, at all intermediate points,
the inequality given in equation \eqref{fully_adhered_cond} holds?' The answer again comes from
the analogy between the normal and shear components of loading, and
is
\begin{align}
q_{B}(x)=q_{A}(x)+\int_{P_{A}}^{P_{B}}p'(x)\frac{\mathrm{d}Q}{\mathrm{d}P}\mathrm{d}P\text{,}
\end{align}
where $p'(x)=\frac{\partial p(x,P)}{\partial P}$. In particular,
point $A$ might be the start of loading. Note that, in this process, we are locking in shear traction and attendant slip displacement.
This is one useful major new result. The second is the observation
that, when a region of frictional slip is developing,  provided that the quantities are interpreted correctly, the
superposition idea of the J\"ager-Ciavarella theorem continues to hold when the normal
load is changing. So if we are at point ($P_{X},Q_{X})$ during steady-state loading
on the trajectory, the superposition (with a minus sign) of
a corrective shear traction from some earlier point ($P_{Y},Q_{Y})$ would
ensure stick over the half width $a_{Y}=a(P_{Y})$. The point $Y$
is located by the imposition of tangential equilibrium, and gives
\begin{align}\label{eq17}
f=\frac{Q_{X}-Q_{Y}}{P_{X}-P_{Y}} \text{.}
\end{align}

It is therefore found by drawing a line of gradient $f$ on the $P,Q$
diagram and finding where the line intersects the loading curve. From
these two principles a complete picture of the evolving stick-slip
pattern, and the shear traction distribution can be tracked out \cite{Barber_2011}, \cite{Hills_2011}.
Note that, after the first cycle of loading the shear traction distribution
changes from the initial one to a steady state distribution, and this
modifies the form of the construction needed. The original papers should
be studied for details, but one thing which emerges is that the size
of the steady state permanent stick region, $a(P_{T})$ is found by drawing
lines of gradient $\pm f$ which are tangential to the envelope of
loading, and these define a point $T$, as shown in Figure \ref{fig:PQSpace}.

Progress was also made on the problem of varying normal and bulk
load alone, so that the slip zones are inherently antisymmetric but equal in
size and magnitude, provided, again, that the contact itself is symmetrical
\cite{Ramesh_2017}. But, there seemed little hope of being
able to extend the idea of the J\"ager-Ciavarella theorem to the case where both shear force and large
bulk tension were present and varying in a complicated way, where large bulk tension means that the direction of slip is opposing at the ends of the contact. Thus, there has been
no development of the method, in this form, for the cases of either
(a) large bulk tension and shear force present, or (b) when the contact
is not symmetrical, either because the indenter itself is not symmetrical
or because there is a moment present.

\section{Solutions on the Basis of Full Stick}

\hspace{0.4cm}All of the solutions presented above hinge on the superposition of
the sliding shear traction distribution and a corrective shear in
the stick region, i.e. of the two mixed boundary conditions to be
enforced that in the slip region is automatically ensured whilst that
in the stick region, which might more generally be thought of as ensuring that any locked-in surface strains are preserved, is found as a secondary consideration.

\begin{figure}[t!]
	\centering
	\includegraphics[scale=0.75, trim= 0 0 0 0, clip]{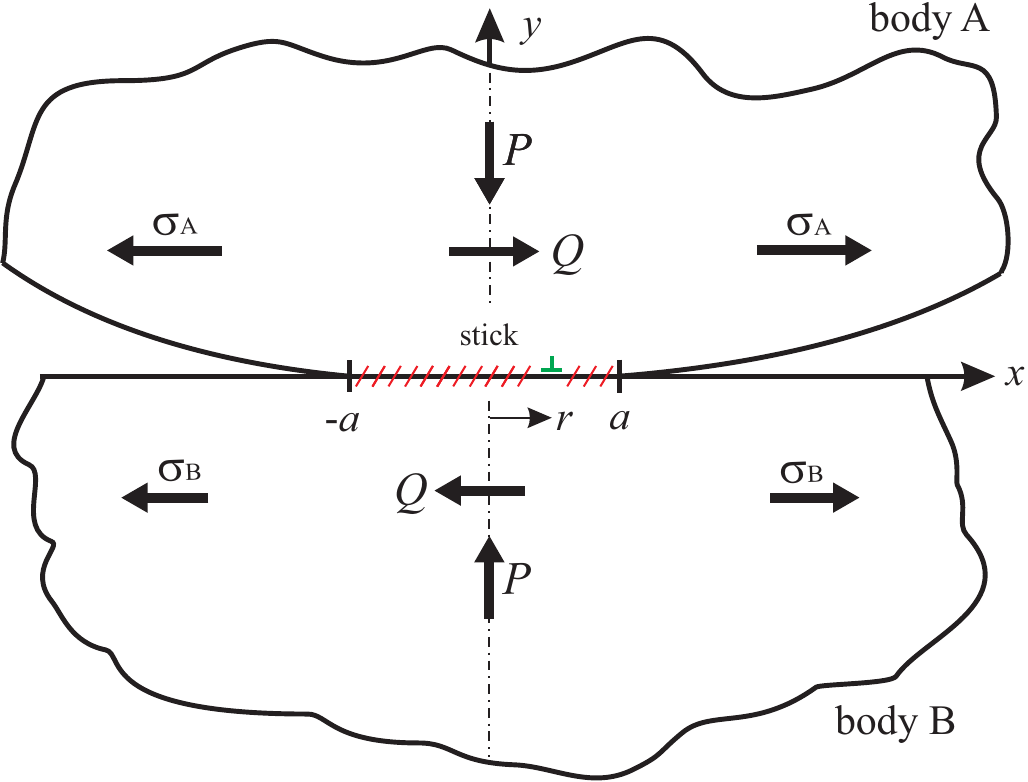}
	\caption{Symmetrical half-plane contact subject to normal load, shear load and bulk tension in a fully adhered condition with the introduction of a single glide dislocation at $x \rightarrow r$.}
	\label{fig:adhered_condition}	
\end{figure}

In this section we explore the possibility of starting off with a
solution which is fully stuck (and which may or may not include a
locked-in displacement), and then adding a perturbation to permit slip
to establish itself. The kernel for this solution, therefore, is not the solution for a line force on a half-plane, but that for an edge dislocation, present on the interface line between
two half-planes in glancing contact and traction-free outside the
region of contact (which we will take as $\left|x\right|<a$), but
bonded together within it, Figure \ref{fig:adhered_condition}. Two break-throughs were needed for this
to be a possibility - one was obtaining the solution for a glide dislocation
in the domain described and the second was devising an inversion procedure
for a Cauchy integral equation imposed over two regions (the two slip
zones), and these are described in the following articles by Hills et al. \cite{Hills_2018} and Moore et al. \cite{Moore_2018}. 

An edge dislocation, glide in character and therefore having Burgers
vector $b_{x}(r),$ installed at point $x=r$, along the line $y=0$
defining the interface between two half-planes adhered together over
the interval {[}$-a\qquad a${]} induces no direct traction along
the interface but a shear traction given by
\begin{align}
q(x)=\frac{E}{2\pi(1-\nu^{2})}\frac{b_{x}\left(r\right)}{r-x}\sqrt{\frac{a^{2}-r^{2}}{a^{2}-x^{2}}}\,\text{,}
\end{align}
under plane strain conditions. This has the anticipated properties; there is a $1/x$-like singular
behaviour as the observation point approaches the dislocation $x \rightarrow r$, and
the traction also becomes singular in an $s^{-1/2}$ manner when the
ends of the interval are approached.

\begin{figure}[t!]
	\centering
	\includegraphics[scale=0.445, trim= 0 0 0 0, clip]{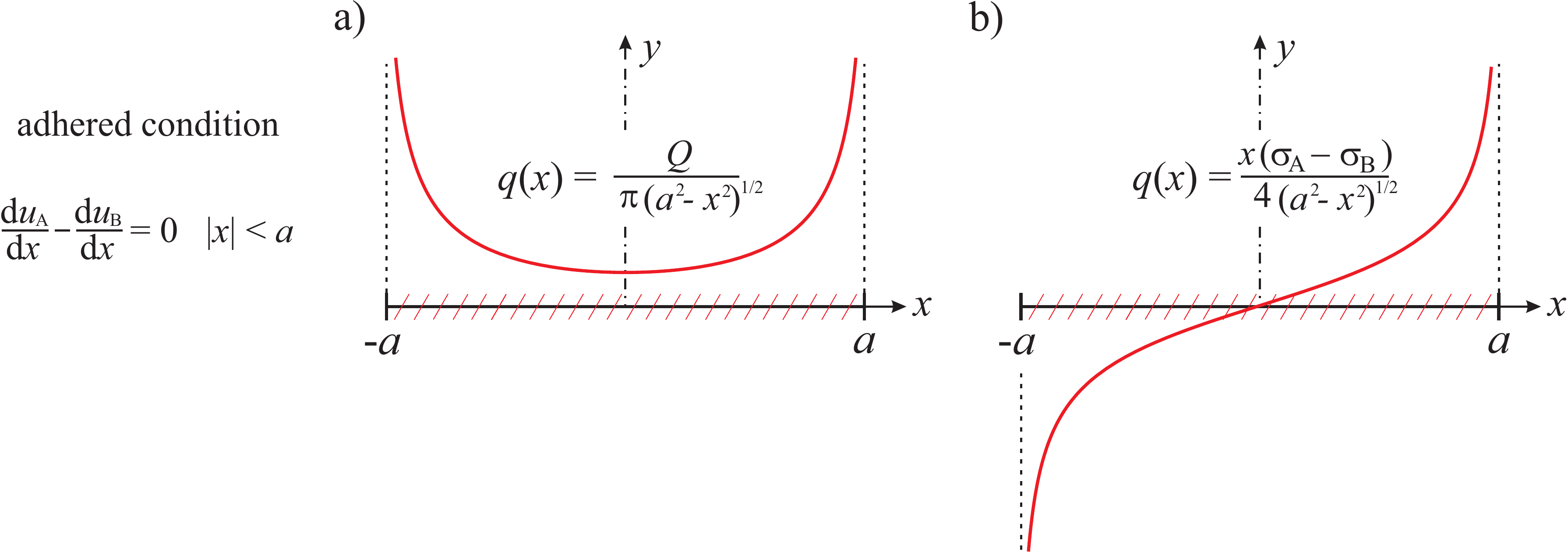}
	\caption{Shear tractions in the fully adhered state due to a) shear force and b) bulk tension.}
	\label{fig:figurefullyadhered}	
\end{figure}

As a educational example, consider the sequentially loaded problem,
where the normal load is applied first and then a shear force, $Q$
is gradually exerted. If the contact is fully stuck the locked in strain difference is, here,
zero throughout, i.e. $\frac{\mathrm{d}u_{\text{A}}}{\mathrm{d}x} - \frac{\mathrm{d}u_{\text{B}}}{\mathrm{d}x} = 0 \, \, \, \, ,|x|<a$. The shear traction induced along the contact, as shown in Figure \ref{fig:figurefullyadhered} a),
$q_{\text{st}}(x)$, will be singular at the edges. Equally, if tensions, $\sigma_A, \sigma_B$, are developed in each body as shown in Figure \ref{fig:adhered_condition} the shear tractions arising will be as shown in Figure \ref{fig:figurefullyadhered} b). Combined, the fully stuck shear tractions are given by
\begin{align}
q_{\text{st}}(x)=\frac{Q}{\pi\sqrt{a^{2}-x^{2}}}+\frac{\sigma x}{4\sqrt{a^{2}-x^{2}}},\quad\left|x\right|<a \text{.}
\end{align}

Suppose that there is a slip zone attached to the left hand edge,
stretching from $[-a\quad-b_{1}]$ and a slip zone attached to the
right hand edge, stretching from $[b_{2}\quad a]$. We therefore
need to distribute glide dislocation over these intervals, and the
dislocation density, $B_{x}=\mathrm{d}b_{b}/\mathrm{d}x$ represents the difference
in surface strains. The corrective traction induced, $q_{\text{corr}}(x)$, is given
by
\begin{align}
q_{\text{corr}}(x)=\frac{E}{2\pi(1-\nu^{2})\sqrt{a^{2}-x^{2}}}\left[\int_{-a}^{-b_{1}}\frac{\sqrt{a^{2}-\xi^{2}}B_{x}\left(\xi\right)d\xi}{\xi-x}+\int_{b_{2}}^{a}\frac{\sqrt{a^{2}-\xi^{2}}B_{x}\left(\xi\right)d\xi}{\xi-x}\right],
\end{align}
for $\left|x\right|<a$.
Note that, within the stick region ($-b_{1}<x<b_{2}$) both integrals
should be interpreted in a regular sense. In the left hand slip region,{[}$-a\quad-b_{1}]$,
the first integral, only, is Cauchy and in the right hand slip region,
{[}$b_{2}\qquad a${]}, the second integral, only, is Cauchy. 

Now,
the sign of slip depends on the relative magnitude of the shear force
and bulk tension; if the shear force is large the slip zones will
be of the same sign whereas if the bulk tension dominates they will
be of opposite sign, but, in either case, we may write
\begin{align}
q_{\text{corr}}(x)+q_{\text{st}}(x)=\pm fp(x),\qquad-a<x<-b_{1},\,b_{2}<x<a \text{.}
\end{align}

The details of the integral equations and their inversion are given
in the original papers cited, and a complementary solution for the
normal contact problem itself, found using dislocations rather than
a line for as the kernel, is given in \cite{Moore_2018_2}.

\section{Further Generalisation and the Steady State}

\begin{figure}[t!]
	\centering
	\includegraphics[scale=0.4, trim= 0 0 0 0, clip]{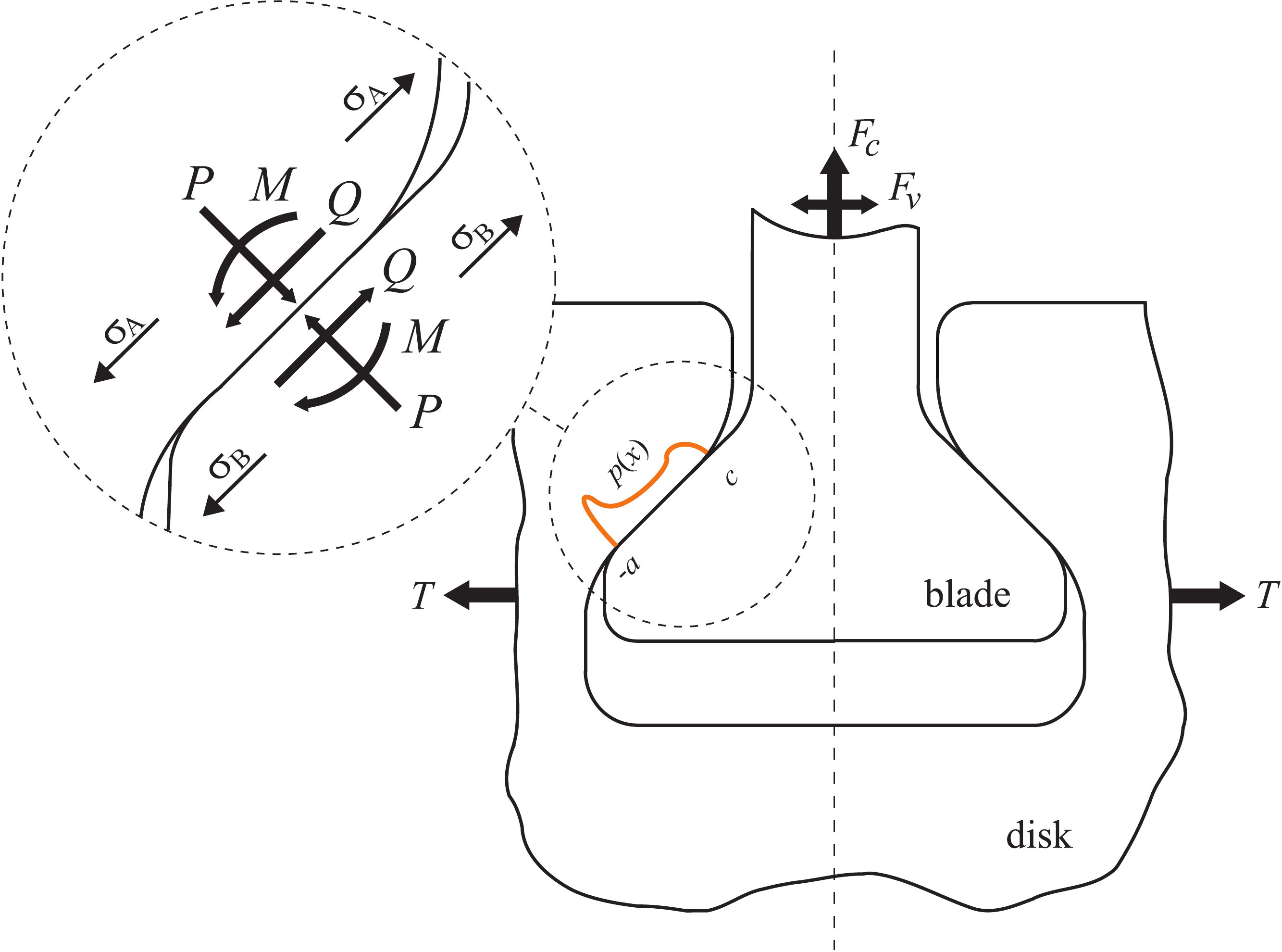}
	\caption{Dovetail root problem with external loads ($F_C, F_V, T$) and contact loads ($P, Q, M, \sigma_A, \sigma_B$).}
	\label{fig:dovetail}	
\end{figure}

\hspace{0.4cm}The formulations described in $\mathsection3$, $\mathsection4$ may be used, in principle,
to track out transient problems on a marching-in-time basis, with
and without a phase shift, although the technique described in $\mathsection3$ is
incapable of handling the simultaneous presence of a shear force and bulk tension. The method outlined in
$\mathsection4$ has greater generality but it requires a numerical
implementation when there is a phase shift, for example. There is
also one feature of many contacts which is absent from the solutions
treated so far, and this is the presence of a moment. A frequently
occurring geometry in practice is, for example, the fan blade dovetail root contact referred to in the introduction, Figure \ref{fig:dovetail}. The blade might be subject to a centrifugal, $F_C$, and vibration load, $F_V$, and the disk experiences an expansion force, $T$. This type of contact may be approximated by a punch having a flat base but rounded corners, and which is therefore capable of sustaining a moment, $M$, unlike a Hertzian contact.
 
Therefore, the most general loading of a half-plane contact is represented by
the four variables $P,Q,M,\sigma$. It is noteworthy that the tendency
to slide is independent of $M,\sigma$, i.e. it depends on the ratio
$Q/P$ alone. Also, the normal load problem depends only on the pair
$P,M$, and is independent of the other two quantities. And, \emph{when
	once the normal problem is solved}, the partial slip problem depends
only on $Q$, $\sigma$, i.e. because the normal and tangential problem are uncoupled if the contacting bodies are elastically similar and the effect of geometric coupling is negligible, these
last two quantities do not feed back into the normal load problem.

Earlier, we gave the condition for no slip where the contact was loaded
by the pair ($P,Q$) alone, equation \eqref{eq17}. Suppose that we have a frictional contact whose instantaneous
half-width is $d$, and we make a small change, $\Delta P$, in normal
load together with a small change in moment, $\Delta M$. If at the same time there are small changes in shear force, $\Delta Q,$ and differential bulk tension, $\Delta\sigma$, the inequality for no slip is 
\begin{equation}\label{inequality}
\left(\frac{\Delta Q}{\Delta P}\pm\frac{\pi}{4}\frac{ d\Delta\sigma}{\Delta P}\right)\left/\left(1\mp\frac{2\Delta M}{d\Delta P}\right)\right.<f,
\end{equation}
where we choose the upper sign when considering the left hand side (LHS)
of the contact and the lower sign when considering the right hand side (RHS) of
the contact. The necessary condition for no slip is that the inequality
holds at \emph{both} sides of the contact.

\begin{figure}[t]
	\centering
	\includegraphics[scale=0.4, trim= 0 0 0 0, clip]{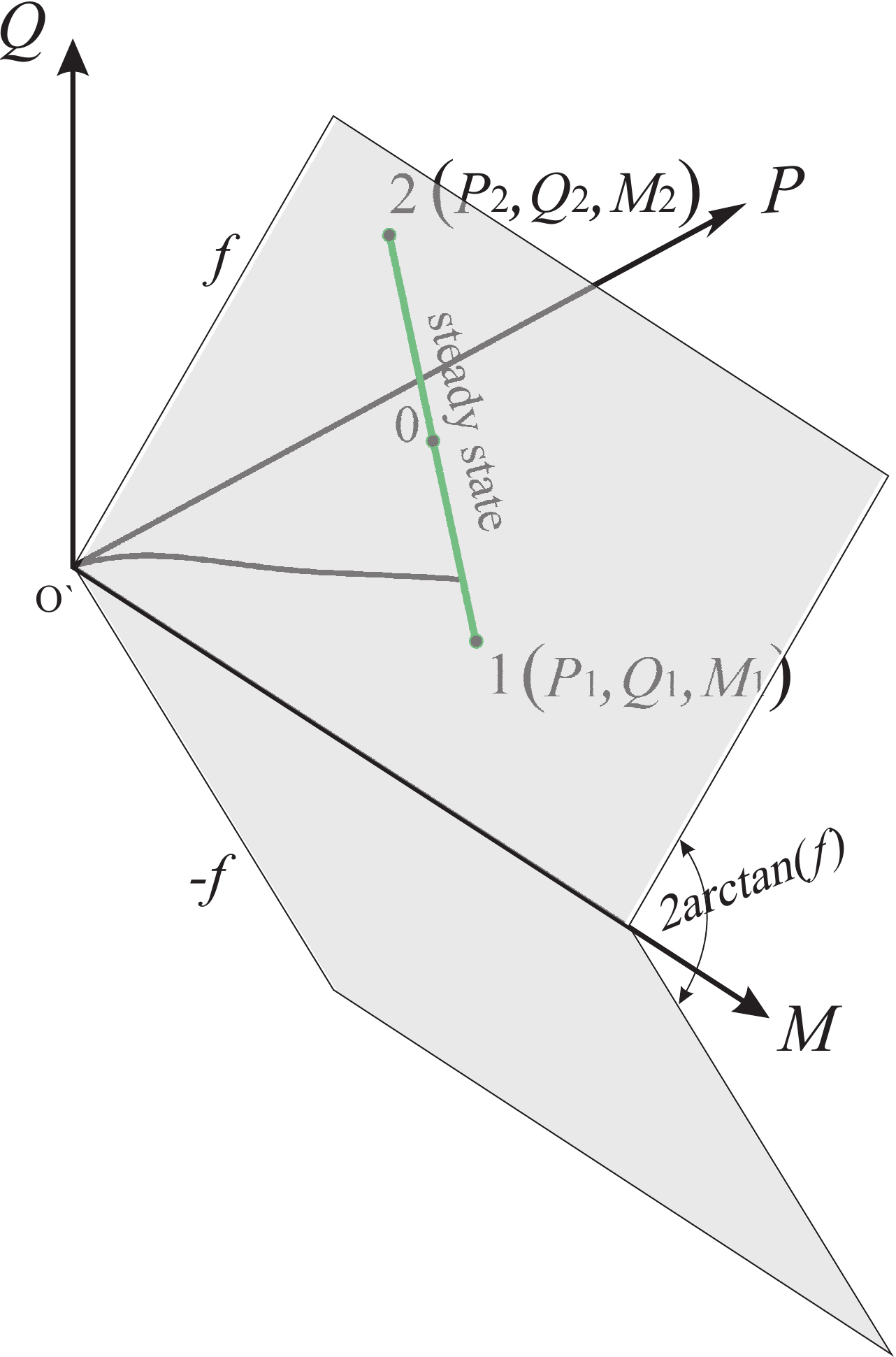}
	\caption{Three-dimensional illustration of $P$-$Q$-$M$ load space for a steady-state partial slip problem, ignoring the bulk tension $\sigma$.}
	\label{fig:PMQsigma_load_space}	
\end{figure}

If we return, again, to the practical problem just mentioned
we note that there are two loads in the system which are substantially
constant, varying only occasionally (the centrifugal load, $F_C$ and expansion force, $T$). Although these
will excite all four components of contact load, they are taken to a mean value
which we denote by the addition of a subscript, $0$. Also, there is a second source of loading on the assembly - vibration
loads, $F_V$ - and these, too exert all four contact loads.
But, because it is a solitary load which excites those changes, they
all occur in phase, which makes the problem rather easier to solve
than the shaft with an eccentric weight referred to in $\mathsection3$,
because there are no phase shifts, and all four quantities change
together. Thus, the problem we have to solve for is one in which the
quantities move along a straight line segment in $P$-$Q$-$M$-$\sigma$ load
space during the steady-state. The condition for the assembly to become locked (even if the
transient movement cause by the primary load violates inequality \eqref{inequality}
is this same one but where the infinitesimal changes are replaced
by the finite ranges of the same quantities in the steady state. When the inequality is
violated, there will be some steady state slip. And, in the problems
cited, there will be usually many tens of thousand of `steady state' cycles for each change in the `mean' load. So, the most general, transient behaviour which we have, so far, studied may not be particularly
relevant, and it is the steady state response (established after a
single cycle) which really matters, and this is relatively easy to
find. It is difficult to visualise the problem in an abstract four-dimensional load space. Therefore, it is valuable to visualise the history of loading in the three-dimensional load space depicted in Figure \ref{fig:PMQsigma_load_space}, ignoring the bulk tension, $\sigma$.  

\begin{figure}[ht]
	\centering
	\includegraphics[scale=0.5, trim= 0 0 0 0, clip]{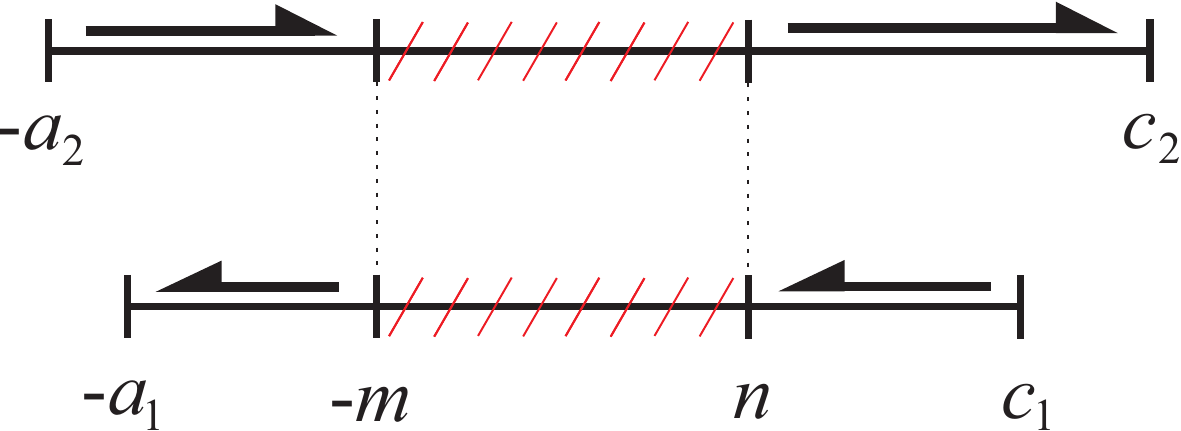}
	\caption{Contact as the ends of loading cycle are approached, including a permanent stick zone.}
	\label{fig:contact_patch}	
\end{figure}

What we need to do is to write down the state of the contact
at the two end points of the steady-state cycle - in the case of the strain/shear traction equation,
immediately \textit{before} the end point is reached, when the slip zones are
at their maximum extent. We will assume that, in the load space described,
the coordinates are given by ($P_{i},Q_{i},M_{i},\sigma_{i}$), where
$i=1,2$, and the first location corresponds to, for example, $P_{1}=P_{0}-\Delta P/2$,
and the second point to, for example, $P_{2}=P_{0}+\Delta P/2$. Figure
\ref{fig:contact_patch} is a schematic of the size and position of the contact at these
extremes, together with an impression of the permanent stick zone.
We shall assume that the bulk load is insufficient to reverse the
direction of slip at the end of the contact where it is subtractive.

Consider, first, the normal loading problem. Although this is not a necessary step, we will assume that the contact
itself is symmetrical in form, so that $\mathrm{d}h/\mathrm{d}x$ is strictly odd, but
permit there also to be a rotation, $\alpha$, which varies with the applied moment. The contact is now
assumed to occupy the interval [$-a_{i}\qquad c_{i}$] and so a
modified form of the integral equation given at the beginning of $\mathsection2$, equation \eqref{SIE}, is needed. At point $i$ in the loading cycle
($i=1,2$), the relative slope of the half-plane surfaces, $\mathrm{d}g_{i}/\mathrm{d}x$,
is given by 
\begin{align}
\frac{E}{1-\nu^{2}}\left[\frac{\mathrm{d}h}{\mathrm{d}x}+\alpha_{i}\right]=\frac{4}{\pi}\int_{-a_{i}}^{c_{i}}\frac{p_{i}(\xi)\,\mathrm{d}\xi}{\xi-x}.\label{ec_07}
\end{align}
and normal and rotational equilibrium are imposed by setting 
\begin{align}
P_{i} & =\int_{-a_{i}}^{c_{i}}p_{i}(x)\,\mathrm{d}x,\label{ec_09}\\
M_{i} & =\int_{-a_{i}}^{c_{i}}p_{i}(x)\,x\,\mathrm{d}x.\label{ec_10}
\end{align}

We turn, now, to the `tangential' integral equation and write that
the difference in surface strains at any location in the surface,
at load state `1' is given by
\begin{align}
\frac{E}{1-\nu^{2}}\,\Delta\varepsilon_{xx,1}=-\frac{4}{\pi}\int_{-a_{1}}^{c_{1}}\frac{fp_{1}(\xi)\mathrm{d}\xi}{\xi-x}+\frac{4}{\pi}\int_{-m}^{n}\frac{q_{1}^{\ast}(\xi)\mathrm{d}\xi}{\xi-x}+\sigma_{1},\label{ec_17}
\end{align}
where permanent stick is present over the interval {[}$-m\qquad n${]}. At load point 2 we write a similar expression, but with the sign of
the overall sliding shear traction reversed, so as to give the correct
sign of shear traction in the end slip regions, as 
\begin{align}
\frac{E}{1-\nu^{2}}\,\Delta\varepsilon_{xx,2}=\frac{4}{\pi}\int_{-a_{2}}^{c_{2}}\frac{fp_{2}(\xi)\mathrm{d}\xi}{\xi-x}+\frac{4}{\pi}\int_{-m}^{n}\frac{q_{2}^{\ast}(\xi)\mathrm{d}\xi}{\xi-x}+\sigma_{2}.\label{ec_18}
\end{align}

The permanent stick zone must be unique, and is that present at each
of these points, in order to ensure continuity of material, so that
the locked-in strain within this region is preserved, i.e.
\begin{align}
\Delta\varepsilon_{xx,1}=\Delta\varepsilon_{xx,2}\,\,,\text{ }-m<x<n.\label{ec_19}
\end{align}

In the permanent stick zone, $-m<x<n$, these equations become 
\begin{align}
-\frac{4\left(1-\nu^{2}\right)}{\pi E}\left(\int_{-a_{1}}^{c_{1}}\frac{fp_{1}(\xi)\mathrm{d}\xi}{\xi-x}+\int_{-a_{2}}^{c_{2}}\frac{fp_{2}(\xi)\mathrm{d}\xi}{\xi-x}\right)-\Delta\sigma=\frac{4\left(1-\nu^{2}\right)}{\pi E}\int_{-m}^{n}\frac{\left[q_{2}^{\ast}-q_{1}^{\ast}\right]\mathrm{d}\xi}{\xi-x},\label{ec_20}
\end{align}
where $\Delta\sigma=\sigma_{2}-\sigma_{1}$. Since $-a_{2}<-a_{1}<-m<x<n<c_{1}<c_{2},$ use may be made of
the `normal' solution, in a manner similar to that use by J\"ager and
Ciavarella, and we may write
\begin{align}
\frac{8Ef}{1-\nu^{2}}\frac{\mathrm{d}h}{\mathrm{d}x}+\frac{8Ef}{1-\nu^{2}}\alpha_{0}-\Delta\sigma=\frac{4}{\pi}\int_{-m}^{n}\frac{\left[q_{2}^{\ast}-q_{1}^{\ast}\right](\xi)\mathrm{d}\xi}{\xi-x},\text{ }-m<x<n,\label{ec_21}
\end{align}
where $\alpha_{0}=\frac{\alpha_{1}+\alpha_{2}}{2}$.
Note that the left hand side consists of only two terms - the gradient
of the surface as it appears in the `normal' integral equation, and
a constant. This means that solution is very straightforward, and
the original papers by Andresen et al. \cite{Andresen_2019_2} and \cite{Andresen_2019_3} should be consulted for
full details. The only additional elements needed are the range of shear force,
$\Delta Q$, which is given by 
\begin{align}
\Delta Q=Q_{2}-Q_{1}=2fP_{0}+\Delta Q^{\ast},\label{ec_25}
\end{align}
where $P_{0}=\frac{P_{1}+P_{2}}{2}$ is the mean normal load, and 
\begin{align}
\Delta Q^{\ast}=\int_{-m}^{n}[q_{2}^{\ast}-q_{1}^{\ast}](x)\mathrm{d}x.\label{ec_26}
\end{align}

\section{Conclusions}

\hspace{0.4cm}We have reviewed the current state of understanding of partial slip
half-plane contact problems, starting from the first steps made by
Cattaneo over eighty years ago. The concepts given have not been attached
to a particular geometry although inevitably the more complicated the indenter
shape the more taxing the algebra is when it comes to implementation.

A full description of the partial slip solution means that we must
know, first, the evolving stick pattern during the load cycle. The
locked in shear distribution within this permanent stick zone has
no practical bearing on the solution, and this means that, in periodic
problems the transient loading from an unloaded state, which affects
precisely just this locked in shear (and attendant slip displacement)
does not have any practical effect on the steady state solution. A
complete solution requires other information - in particular the slip
displacement - but this can be evaluated when once the stick zone is
established, by various means. And, when once that is done, other
derivative information - the pointwise frictional energy expenditure,
total frictional damping energy, surface stress etc. can all be found
quite easily.

The paper distinguishes clearly between tracking out problems in a
marching-in-time sense, including transient conditions, and separates
these from the simpler but often more important steady state response.
It looks at alternative methods of formulation based on perturbations
of the full-stick solution, using dislocations as the kernel. The
question of asymptotic representations of the slip zone conditions,
which are of practical interest in matching test conditions, is specifically
excluded.

\section*{Acknowledgements}
\hspace{0.4cm}This project has received funding from the European Union's Horizon 2020 research and innovation programme under the Marie Sklodowska-Curie agreement No 721865. David Hills thanks Rolls-Royce plc and the EPSRC for the support under the Prosperity Partnership Grant 'Cornerstone: Mechanical Engineering Science to Enable Aero Propulsion Futures', Grant Ref: EP/R004951/1.

\bibliography{Contribs_References_Hendrik_Jan2019}

\clearpage{}

\end{document}